\documentclass{article}

\usepackage{color}
\usepackage{booktabs}
\usepackage{array}
\usepackage{tabularray}
\usepackage{authblk}
\usepackage{pifont}
\usepackage{amsmath,amssymb,amsfonts}
\usepackage{algorithmic}
\usepackage{graphicx}
\usepackage{multirow}
\usepackage{svg}
\usepackage{textcomp}
\usepackage{xcolor}
\usepackage{diagbox}
\usepackage{adjustbox}
\usepackage{hyperref}
\hypersetup{
    colorlinks,
    citecolor=black,
    filecolor=black,
    linkcolor=black,
    urlcolor=black
}
\usepackage{xcolor,colortbl}
\usepackage{tcolorbox}
\usepackage{rotating}
\usepackage{booktabs}
\usepackage[square,numbers, sort&compress]{natbib}
\usepackage[a4paper]{geometry}
\newgeometry{
    left=2.2cm,    
    right=2.2cm    
}

\definecolor{Gray}{gray}{0.85}
\definecolor{LightCyan}{rgb}{0.88,1,1}
\definecolor{Red}{rgb}{1.0,0.15,0.0}
\definecolor{Orange}{rgb}{0.8, 0.33, 0.0}
\definecolor{LightOrange}{rgb}{1.0,0.87,0.0}
\definecolor{gray}{rgb}{0.75, 0.75, 0.75}
\definecolor{gray_dark}{rgb}{0.70, 0.70, 0.70}
\definecolor{gray_light}{rgb}{0.86, 0.86, 0.86}
\newcolumntype{a}{>{\columncolor{Gray}}c}
\newcolumntype{b}{>{\columncolor{white}}c}

\begin{document}

\title{Identification of Technical Design Constraints and Considerations for Transmission Grid Expansion Planning Projects}

\author[1]{Giacomo Bastianel}
\author[2]{Clement Hardy}
\author[2]{Nils Charels}
\author[1]{Dirk {Van Hertem}}
\author[1]{Hakan Ergun}

\affil[1]{{Department of Electrical Engineering, KU Leuven, Leuven, Belgium, and Energy Transmission Competence Hub (Etch) - EnergyVille, Genk, Belgium} \newline}
\affil[2]{{Elia Transmission Belgium, Brussels, Belgium}}

\date{}

\maketitle

\begin{abstract}
The large-scale deployment of renewable energy sources, particularly offshore wind, requires large-scale transmission grid expansion projects to transmit the produced low-carbon power to the main demand centers. However, the planning and design of such complex projects currently lack a transparent and systematic process that system operators can follow when considering such investments in their grids. This paper identifies and classifies the main technical design constraints and considerations relevant to the planning of transmission grid expansion projects, and more specifically, electrical energy hubs. Seven key areas of interest are identified focusing mainly on European projects, namely \textit{network integration}, \textit{HVDC technologies}, \textit{costs} (CAPEX, OPEX, and space requirements), \textit{electricity market design}, \textit{future proofness and modular expandability}, \textit{reliability-availability-maintainability}, and \textit{sustainability}. Each area of interest is analyzed in terms of its technical and operational relevance, with technical design constraints and considerations derived from such analysis. In addition, a hierarchical classification of the identified constraints and considerations (and therefore areas of interest) is introduced, distinguishing them between three criticality classes, namely \textit{hard constraints}, \textit{main drivers}, and \textit{key considerations}. The dependencies between the different areas are discussed mainly for the European context, too. Therefore, this work provides system operators and policymakers with a structured basis to support a transparent planning methodology with clear decision hierarchies for investments in transmission grid expansion projects.
\end{abstract}

Keywords: Electrical energy hubs, HVDC technologies, Multi-terminal HVDC grids , Network integration , Technical design constraints , Technical design considerations

\section{Introduction} \label{sec:intro}
In its decarbonization plans, the European Union aims to install 450 GW of offshore wind capacity in its territory by 2050~\cite{fitfor55}. Furthermore, the Esbjerg (2022)~\cite{Esbjerg_declaration}, Ostend (2023)~\cite{Ostend} and Hamburg (2026)~\cite{Hamburg_declaration} North Sea summits pledge to install 300 GW offshore wind capacity by 2050 in the North Sea area. An efficient transmission system design is needed to transport such amounts of power to shore in an affordable way. In this regard, High Voltage Direct Current (HVDC) links proved to be the most cost-effective means of transmitting power over long distances~\cite{DirkHVDCgrids2016,Hardy_powertech}. So far, HVDC links have mainly been Point-to-Point (PtP), connecting two converters within the same synchronous area, e.g. the ALEGrO link between Belgium and Germany~\cite{ALEGRO}, in two different asynchronous areas, e.g. the Nemo link between Belgium and the UK~\cite{NEMO}, or to radially deliver electricity generated by a wind farm to the onshore grid, e.g. in BorWin 1 in the Netherlands~\cite{BorWin1}. Recently, the idea of using hybrid interconnectors linking two control areas with a wind farm in between them has been explored, for example, for the Nautilus project between Belgium and the UK~\cite{Nautilus} and the HansaLink between Germany and the UK~\cite{HansaLink}. With extended offshore wind projects, e.g. as the Dogger Bank wind farm in the UK~\cite{Dogger_bank}, and growing HVDC grid projects~\cite{Bastianel_AEIT,Terna}, the aggregation of multi-GW power capacity and interconnection to different countries has been introduced in~\cite{Meijden_Tennet} and recently proposed in \textit{offshore energy hubs}~\cite{Luth} and \textit{energy islands} projects from the Belgian (Elia)~\cite{Eliaenergyisland} and Danish (Energinet)~\cite{DKislands} transmission system operators (TSOs), supporting the need for interconnections currently existing in the European power system~\cite{JRC_2025}. Furthermore, the North Sea Wind Power Hub project proposed to create a network of islands with its hub-and-spoke concept~\cite{NSWPH}. 

As these projects can be proposed onshore with other types of renewable energy sources (RES) with the same technical challenges, we focus on offshore electrical energy hubs (EEHs) and their functional specifications as discussed in~\cite{Definitions_paper}. 
An EEH is defined as a ``\textit{multifunctional node in a power system, managed separately from the main existing control areas, aggregating local GW-size electrical generation capacity. It is characterized by a high density of electrical equipment and has multiple interconnections to existing control areas}''. Given the nature of such projects, their efficient planning is crucial to maximizing both the power generation and transmission to existing control areas while guaranteeing safe and reliable operations of the entire grid. To reach these goals, this paper identifies and discusses the main technical constraints to be considered by transmission system operators (TSOs) and asset owners during the planning stage of EEHs. Note that while the paper is mainly focused on European EEH projects and therefore European-specific regulatory requirements, we indicate which technical constraints and considerations are universally valid, and which are linked to the regulations of the area where the EEH is built. After this introduction, a general overview of the identified technical constraints and the motivation behind this work is presented in Section~\ref{sec:Motivation}. Section~\ref{sec:constraints} discusses each constraint related to the planning of EEHs and the areas of interest to which they relate. Moreover, Section~\ref{sec:discussion} classifies the constraints following a hierarchy based on their criticality for the project. They are divided into three classes, namely \textit{hard constraints}, \textit{main drivers}, and \textit{key considerations}. Finally, Section~\ref{sec:conclusion} concludes the paper and describes possible future work related to the planning of EEHs.
\section{Motivation and contribution} \label{sec:Motivation}

The EEH projects foreseen in the North Sea~\cite{Eliaenergyisland,DKislands} combine the aggregation of multi-GW offshore wind capacity with several \textit{interconnections} to existing \textit{control areas}\footnotemark[1]. Given the expected multi-GW size of these projects, there is a strong need to efficiently integrate them into the existing power system, as they can positively benefit the overall socioeconomic welfare of the control area where they are installed by providing large amounts of low-carbon energy.

\begin{figure}
    \centering
    \includegraphics[width=0.65\linewidth]{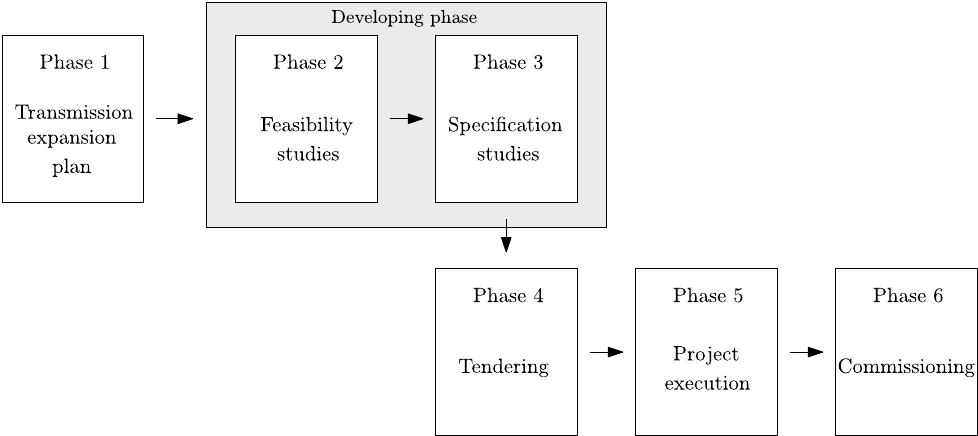}
    \caption{Six main phases of the transmission expansion planning process~\cite{Stu_for_plan_HVDC}. This work identifies and discusses relevant technical constraints for investments in electrical energy hub projects, and focuses on the first \textit{transmission expansion plan} and second \textit{feasibility studies} phases.}
    \label{fig:phases}
\end{figure}

For this reason, this work aims to extend the findings from the ``IEEE Technical report PES-TR86 - Studies for Planning HVDC''~\cite{Stu_for_plan_HVDC}, which focuses on studies for planning new HVDC links to be added to existing power systems. The technical report~\cite{Stu_for_plan_HVDC} distinguishes six main phases for the transmission development process, ranging from techno-economic analyses in the early phases of the planning process to detailed engineering decisions during the actual project realization. The six phases are displayed for clarity in Figure~\ref{fig:phases}, with this paper focusing on the first two.

The \textit{first phase} establishes a transmission expansion plan where high-level capacity needs are identified using market simulations. Furthermore, bottlenecks in the existing grid are identified, usually relying on optimal power flow (OPF) simulations. Several investment options for grid expansion are determined based on the identified bottlenecks. Finally, these grid expansion options are compared using Cost-Benefit Analysis (CBA) methodologies based on several economic, technical, social, and environmental criteria, such as, e.g., the project of common and mutual interest~\cite{PCI_PMI} in the ENTSO-e CBA~\cite{ENTSO_E_CBA}. The output of the first stage is usually the required capacity, location, and technology of the new investments.
 
 The \textit{second phase} consists of conducting feasibility studies for the chosen investment options, where the detailed technical parameters of the investments are identified. Examples are electrical parameters (voltage and current rating, number of circuits needed, impedance values, etc.), high-level substation design, protection requirements, and high-level control requirements to achieve grid connection compliance.

Due to their novelty and a lack of a coordinated master plan for grid expansion (in Europe)~\cite{IEA_2023}, no clear guidelines regarding the planning of EEHs have been discussed in the literature. Particularly, a hierarchy linking the main technical constraints behind the initial design choices for EEHs has never been proposed. Identifying such technical constraints can help transmission asset owners to streamline the EEHs' planning process and guarantee transparency in the decisions regarding their final design. In addition, these technical constraints can represent the first step toward a master plan for transmission expansion planning through EEHs connecting several countries, as proposed, e.g., in the North Sea Wind Power Hub project~\cite{NSWPH} and in~\cite{Bastianel_AEIT}. For these reasons, the contributions behind this work are:
\begin{itemize}
    \item Identification of the main areas of interest to be considered by system operators when evaluating investments in EEHs.
    \item Identification of the main technical constraints and considerations related to EEH's planning and classification into three categories, namely \textit{hard constraints}, \textit{main drivers}, and \textit{key considerations}. Based on these three categories, relations between different constraints are drawn, while they are ordered hierarchically according to their importance.
\end{itemize}

\footnotetext[1]{{}{The ``\textit{interconnection}'' and ``\textit{control area}'' terms used in this paper refer to the definitions from the UCTE glossary~\cite{Glossary}.}}
\section{Identified technical constraints} \label{sec:constraints}

\begin{figure}
    \centering
    \includegraphics[width=0.55\linewidth]{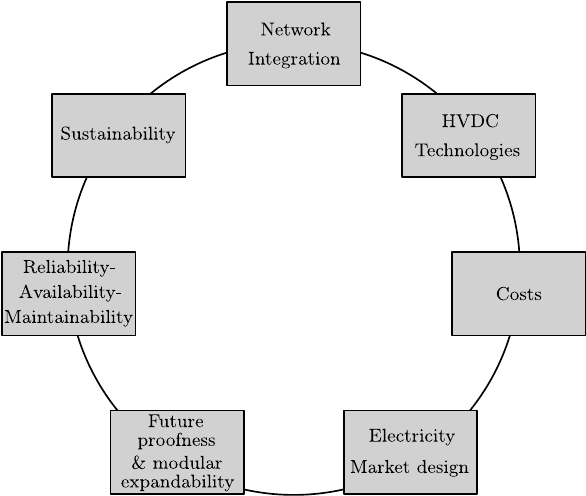}
    \caption{Main areas of interest of the identified constraints related to the planning of electrical energy hub projects in the power system.}
    \label{fig:overview}
\end{figure}

Seven main areas of interest related to the planning of EEHs are identified in the paper and shown in Figure~\ref{fig:overview}. These areas feature purely technical topics such as \textit{network integration}, \textit{HVDC technologies}, \textit{future proofness \& modular expandability}, and \textit{reliability-availability-maintainability}, techno-economical ones such as \textit{costs} (CAPEX, OPEX, and space requirements) and \textit{electricity market design}, and one societal, \textit{sustainability}. While EEHs are transmission technology agnostic, we include HVDC technologies among the main areas of interest. These technologies are widely adopted for transmitting power over long distances, and their use in multi-terminal and meshed grids connecting several EEHs is expected to increase~\cite{ONDP}, although several technical challenges remain to be addressed. Therefore, we consider it important to devote particular attention to HVDC technologies.

\subsection{Network integration}\label{sec:NOS}
The section is divided into two parts, namely \textit{network operational security} and \textit{grid reinforcements and congestion management of the existing grid}. The first includes the power system as a whole, while the latter discusses solutions to reduce grid congestion and guarantee efficient power transmission.

\subsubsection{Network operational security} \label{ch:NOS}
EEHs include several GW of RES installed power capacity, of which the electricity is transmitted to control areas on the onshore grid. Given their considerable size, new investments in EEHs should not jeopardize the \textit{operational security} of the power system to which they are connected, which is defined as ``\textit{the transmission system's capability to retain a normal state or to return to a normal state as soon and as close as possible, and is characterized by thermal limits, voltage constraints, short-circuit current, frequency limits, and stability limits}'' by ENTSO-E's network code on operational security~\cite{Operational_Security}. To that end, TSOs apply the N-1 security criterion, which protects the power system from operation conditions leading to limit violations of the grid's technical constraints as a result of an outage of a single grid element. To ensure the N-1 security criterion, TSOs apply \textit{remedial actions} (RAs), defined as ``\textit{..any measure applied by a TSO to maintain operational security. In particular, remedial actions serve to fulfill the (N-1) criterion and to maintain operational security limits}.''~\cite{Operational_Security}. 
Note that while this Section is focused on the practical example of the frequency regulation mechanisms in continental Europe, the same reasoning can be applied to other areas in the world.

\begin{figure}
    \centering
    \includegraphics[width= 0.65\linewidth]{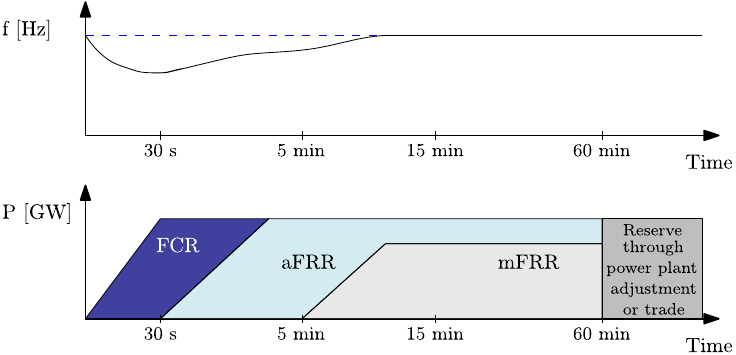}
    \caption{Division of frequency regulation with exemplary frequency curve (top) and power type responsibilities (bottom)~\cite{Frequency}.}
    \label{fig:frequency_restoration}
\end{figure}

Therefore, in case of a loss of infeed in continental Europe's power grid, several mechanisms ensure that the frequency is always within a given safe operational range. The frequency regulation mechanisms are divided into Frequency Containment Reserves (FCR)~\cite{FCR}, automatic Frequency Restoration Reserves (aFRR)~\cite{aFRR}, and manual Frequency Restoration Reserves (mFRR)~\cite{mFRR}. The timescale of the three mechanisms is shown in Figure~\ref{fig:frequency_restoration}.

FCR are European-wide reserves with a maximum FCR contracted capacity of 3~GW at the continental synchronous area level. With the use of Under Frequency Control Schemes (UFCS) for generators if the loss of power is higher than that limit~\cite{entso_e_task_force}\footnotemark[2]. Contrarily, in continental Europe, aFRR and mFRR mechanisms needed to be activated by the TSO of the country where the contingency takes place. Therefore, the dimensioning of the aFRR and mFRR needs to be computed by each TSO for their power system. Consequently, aFRR/mFRR reserves are dimensioned according to the effect of possible contingencies happening in a TSO's grid. For example, expectations for the mFRR and aFRR reserves to be adopted in Belgium are dependent on the expected increase in forecast error linked to (the ramping of) renewable energy sources generation, as indicated by the ``Adequacy and flexibility study for Belgium, 2026-2036''~\cite{Elia_2026_2036}: ``\textit{Fast flexibility: the need for flexibility that can react within 15 minutes of real time – to cover prediction errors or generation and transmission (HVDC) asset outages – is expected to double, exceeding 3 GW in 2036.}''

In this regard, platforms such as the Platform for the International Coordination of Automated Frequency Restoration and Stable System Operation (PICASSO)~\cite{Picasso} and the Manually Activated Reserves Initiative (MARI)~\cite{MARI} have been established. They aim to create platforms for the exchange of balancing energy from frequency restoration reserves with automatic activation, or aFRR-Platform (PICASSO), and future development and operation of manual activation reserves, or mFRR-platform (MARI). The main rationale behind these platforms is to increase the economic efficiency of contracting aFRR and mFRR through cross-border competition. 

As EEHs aggregate multi-GW generation capacity and transmit the generated electricity via several interconnections to existing main control areas, several network elements can create deep frequency deviations in case of faults. These deviations can lead to a further loss of infeed as other generation assets might disconnect due to their under-frequency protection. Consequently, both the conceptual planning of EEHs and their real-time operation are constrained by the FCR requirement and the maximum allowable frequency deviation in the network. A risk assessment approach computing the available transmission capacity at any point in time by the EEH is employed by system operators.

It should be noted that the reserve dimensioning values discussed above, including the projected fast-flexibility requirement of more than 3 GW, are subject to ongoing review by ENTSO-E and the relevant TSOs as the grid conditions, renewable penetration, and regulatory frameworks evolve. Consequently, these values should be regarded as representative of the current planning assumptions at the time of writing rather than fixed design criteria. When assessing the technical feasibility of an EEH, the applicable reserve requirements and operational security criteria should therefore always be verified against the latest version of the relevant ENTSO-E network codes, methodologies, and TSO requirements in force at the time of the feasibility study.

\begin{tcolorbox}
[width=\linewidth, sharp corners=all, colback= white!90!black]
\textbf{Constraint}: 
The grid must always be operated to avoid the presence of any contingency with an instantaneous impact larger than the dimensioning incident for the given power system (e.g. the European maximum FCR level), and with a lasting impact larger than the frequency restoration reserves in any control area.

\end{tcolorbox}

\subsubsection{Congestion management and reinforcements of the existing grid} \label{sec:congestion}
 
In the absence of enough transmission capacity, these power flows create \textit{physical congestion} in the power grid, which is defined as ``\textit{any network situation where forecasted or realized power flows violate the thermal limits of the elements of the grid and voltage stability or the angle stability limits of the power system}''~\cite{capacity_and_congestion}. Physical congestion can worsen with new considerable loads added to the grid, such as data centers, which typically desire to be connected to the power grid within a short period of time, but create further congestion due to their large demand~\cite{IEA_data_centers}. A common, yet expensive, measure used by TSOs to alleviate congestion is the redispatch of generators/loads~\cite{EU_congestion}. In Germany, for example, congestion costs have approximately counted for 2.774 bn€ in 2024 alone~\cite{DE_COSTS,SMARD}. This issue goes beyond the European power grid, with, for example, congestion costs of 13.5-8.33 bn\$ in 2023 and 2024~\cite{Congestion_US_24,Congestion_US_25} for all the US Independent System Operators. Therefore, to accommodate a deeper penetration of RES generation in the power grid while maintaining the system's adequacy and security, remedial actions such as the tap changing of Phase Shifting Transformers (PSTs), Optimal Transmission Switching (OTS)~\cite{Fisher2008,Hedman2009,Hedman_2008}, and Busbar Splitting (BS)~\cite{Heidarifar2021,Morsy2022,Jack,Bastianel_2025} are getting increasing attention to optimize the grid topology at any point in time, and finding relevant areas in the grid to be optimized~\cite{Bastianel_2025_PSCC}. Such RAs fall within the extended concept of ``Grid-enhancing technologies'' (GET), also including flexible AC transmission systems, dynamic line rating and HVDC interconnectors~\cite{MIRZAPOUR2024110304}. The main idea behind such GET technologies is to maximize the transmission capacity in the grid by using it more dynamically~\cite{EPRI,WATT}, while waiting for effective grid expansion projects. In this sense, Transmission Network Expansion Planning problems (TNEP) identify the optimal grid reinforcements and extensions for a power system while including RES generation capacity~\cite{Lumbreras,TNEP_Jay}. In both onshore and offshore grid expansion plans, EEHs are pivotal in aggregating and transmitting considerable amounts of electricity throughout the power grid. 

On the one hand, the first offshore grid expansions were built in the context of a greenfield approach, aiming to maximize the RES generation of (mainly) single offshore wind farms connected with a point-to-point link to the onshore grid~\cite{Alpha_Ventus}. With offshore EEHs, the congestion of the part of the onshore grid closest to the shore becomes a compelling issue. Specifically, onshore grid reinforcements are needed to guarantee redundancy and reliability in the power transmission. In the future, with many EEH projects being proposed in areas with a limited extension, such as the North Sea \cite{ELIA_PEI,DanishWebsite}, thorough planning of each investment and its interconnections to other existing/proposed projects is key in maximizing the generation and transmission capacities among countries. Building transmission infrastructure in such a coordinated way results as well in a minimization of grid congestion.

On the other hand, onshore EEH projects are mostly built in the context of a brownfield approach, where an efficient integration in the existing grid infrastructure is crucial. As adding GWs of capacity (and related transmission infrastructure) to the generation fleet of a power system can create undesired congestion, its resilience and redundancy need to be evaluated for a large set of operating conditions~\cite{Lin_Resilience}, e.g. with high demand hours or in case of contingencies, while taking into account all the technical~\cite{IEA_2025_constraints} and societal issues~\cite{NEUKIRCH_Grid_acceptance} which might delay such projects.

For both cases mentioned above, the available hosting capacity of the existing transmission grid needs to be quantified before considering possible candidates for further transmission expansions, similar to the methods currently used in distribution grids~\cite{Hosting_capacity}. Then, new potential transmission corridors should be identified and optimized. As advocated by~\cite{EC_CBA}, among its Key Performance Indicators (KPIs), the evaluation of candidates should also include grid congestion and RES curtailment based on detailed nodal grid models going beyond the current zonal methodology used in current studies~\cite{TYNDP,ENTSO_E_CBA}. 

Note that the aforementioned reasoning about a more flexible utilization of the transmission grid and efficient onshore and offshore expansions are translated into several initiatives, such as the ``Non-costly remedial action optimization''~\cite{nrao} or ``Regional operational security coordination''~\cite{ROSC} initiative in the Core capacity calculation region in Europe, and are relevant in the future of the European power system depicted in the EU Grids package from December 2025~\cite{EU_grids_package}, and in offshore grids such as the one expected to be built in the North Sea~\cite{ONDP,OTC_II,OTC_III}.

\begin{tcolorbox}
[width=\linewidth, sharp corners=all, colback= white!90!black]
\textbf{Consideration}: Topological remedial actions and efficient grid expansion plans have the potential to reduce grid congestion. EEHs are efficiently integrated into the grid by i) selecting appropriate connection points for new investments and ii) appropriately selecting grid reinforcement projects. 
\end{tcolorbox}
\footnotetext[2]{{}{The maximum allowable loss of infeed is defined as 1.8~GW for Great Britain~\cite{SSQS} and 1.4~GW in the Nordic power grid~\cite{Nordics}.}}

\subsection{HVDC technologies} \label{sec:HVDC}
EEHs aim to aggregate and transmit large amounts of power over long distances. Since HVDC transmission systems are cheaper than AC technologies over long distances (500-800 km for onshore connections, and $\sim$100 km for offshore cables) ~\cite{DirkHVDCgrids2016,hardy_techno-economic_2021}, and guarantee enhanced power flow controllability and frequency support between asynchronous zones~\cite{hertem_multi-terminal_2010}, this section briefly describes the building blocks of this technology, and refers to relevant literature on them.

As introduced in Section~\ref{sec:NOS}, the maximum FCR level is the main technical constraint for the network operational security of an EEH. As a result, the required number and type of HVDC cables and HVDC converters need to be selected such that the outage of a single element does not lead to a larger loss of infeed (LoI) than the allowable limit. Furthermore, the protection strategy of the EEH, e.g. selectivity~\cite{Ss_future_hvdc_grids}, needs to complement the chosen grid design to fulfill frequency regulation constraints, i.e., the maximum LoI.

\subsubsection{HVDC cables configurations}
\begin{figure}
    \centering
\includegraphics[width=0.45\linewidth]{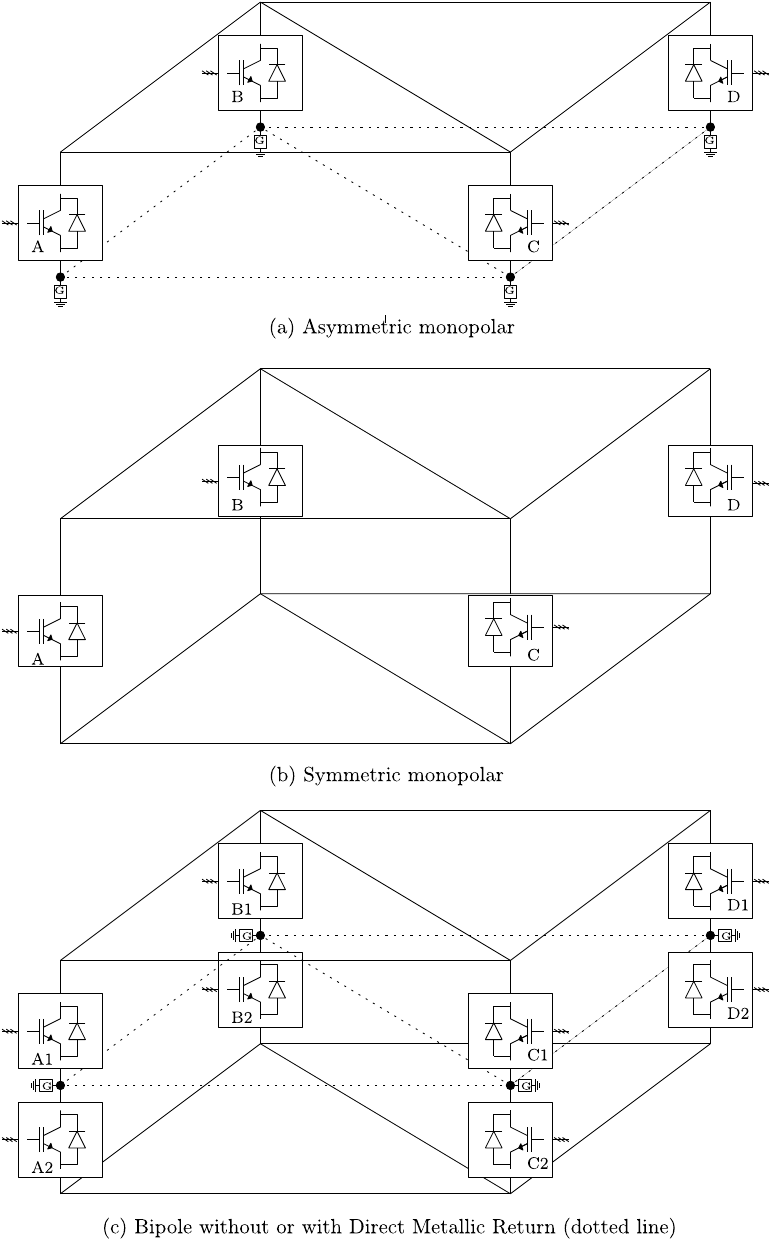}
    \caption{Configurations and grounding for HVDC grids: (a) asymmetric monopolar, (b) symmetric monopolar, and (c) bipolar~\cite{Ss_future_hvdc_grids}.}
    \label{fig:cable_topologies}
\end{figure}

Three main configurations of HVDC systems exist and influence the EEH's DC substation layout. Each of them includes one or two poles, with a potential ground connection and metallic return, as shown in Figure~\ref{fig:cable_topologies}.
They can be classified as~\cite{Ss_future_hvdc_grids}:
\begin{itemize}
    \item{\textbf{Asymmetric monopole}: the current flows through the pole at high DC voltage \texorpdfstring{$V_{dc}$}{} and returns via the metallic return, which is low-impedance grounded. The configuration is shown in Figure~\ref{fig:cable_topologies}~(a).}
    \item{\textbf{Symmetric monopole}: both poles are at high DC voltage with opposite signs \texorpdfstring{$\pm V_{dc}$}{}. The DC side is ungrounded or high-impedance grounded, and the ground connection can be given at multiple points. The configuration is shown in Figure~\ref{fig:cable_topologies}~(b).}
    \item{\textbf{Rigid bipole}: two poles are used at high DC voltage with opposite signs \texorpdfstring{$\pm V_{dc}$}{}. The amount of power that can be transferred is increased compared to the monopoles, but only one DC side current loop exists during normal operation~\cite{HIRSCHING2020106768}. Therefore, the rigid bipole can only operate in symmetrical conditions, and a fault on one of the poles leads to the full cable being unavailable. The configuration is shown in Figure~\ref{fig:cable_topologies}~(c), without the dotted line.}
    \item{\textbf{Full bipole}:
    Two poles are used at high DC voltage with opposite signs \texorpdfstring{$\pm V_{dc}$}{}. The third conductor is low-impedance grounded and is used as a Dedicated Metallic Return (DMR). Differently from the rigid bipole, full bipoles can be operated in an unsymmetrical manner with different powers flowing through the two poles~\cite{Jat2024}, as the DMR guarantees enhanced flexibility. This configuration is currently the standard for offshore HVDC cables set by the Dutch-German TSO TenneT with their 2GW 525 kV plan~\cite{Tennet_2GW}. Note that this configuration is more expensive than the rigid one because of the presence of the DMR. Especially with long cables, the difference in cost can become substantial. For this reason, the benefit brought by a DMR to the system needs to be assessed to select the most cost-effective option. The configuration is shown in Figure~\ref{fig:cable_topologies}~(c).}
\end{itemize}

Note that bipoles can use the ground as a return in case of faults, while they are symmetrical with no ground current during normal operations. In case of fault, the current can continue to flow using the earth as a return through ground return electrodes installed at each end of the line, operating in monopolar mode. Direct ground return is often not allowed by grid codes due to its environmental impact. As a result, the DMR, although more expensive, is the preferred solution~\cite{Direct_ground}. 


\subsubsection{Modular Multilevel Converter types}
\begin{itemize}
    \item \textit{Voltage Source Converter}:
Among the different available converter types, Modular Multilevel Converter (MMC) -Voltage Source Converters (VSC) are the most efficient topology~\cite{Wenig2019_PhD} as:
\begin{itemize}
    \item{They decrease or do not need AC filtering, reducing converter station space requirements.}
    \item{They can control active and reactive power independently at the AC point of common coupling.}
    \item{The modular nature of the Sub-Modules (SMs) in the MMC guarantees more flexible operational features by e.g. using energy storage to reduce fluctuations and AC/DC interactions.}
    \item{Unplanned maintenance actions and downtime are reduced by using redundant SMs to improve the maintenance strategies.}
    \item{Lower losses occur due to the reduced switching frequency of power electronic devices.}
\end{itemize}

The thermal issues created by contingencies that can not be handled by control actions are the most dangerous to the converters. The design of the MMC-VSC converters should deal with the severity of such contingencies, selecting between Half-Bridge (HB), Full-Bridge (FB), or extended topologies. The reader is referred to~\cite{Wenig2019_PhD} for a thorough description of each topology. 

\item \textit{Line Commutated Converter}: Line Commutated Converters (LCC) HVDCs use particular transformers with on-load tap changers and different available winding configurations~\cite{LCC}.
LCC HVDCs have high current capabilities and are suitable for transferring high ($>$ 2GW) amounts of power with good reliability and low maintenance, but they are active consumers of reactive power, due to their inductive behavior. In addition, these converters required advanced filtering and high short-circuit powers. 
\end{itemize}

Therefore, large reactive power compensation installations often need to be added to LCC converter stations. Contrarily, VSC HVDCs are to inject and absorb reactive power into the grid, which allows active voltage control of the transmission grid. In addition, LCC HVDCs need extended AC filtering. These reasons make VSC HVDCs more suitable for offshore EEHs, but LCC HVDCs links might still be used with onshore systems where a high amount of power needs to be transmitted radially from a remote location and a low amount of interconnections.  

\subsection{Protection strategies} \label{subsec:prot_strat}
\begin{figure}
    \centering
    \includegraphics[width=0.6\linewidth]{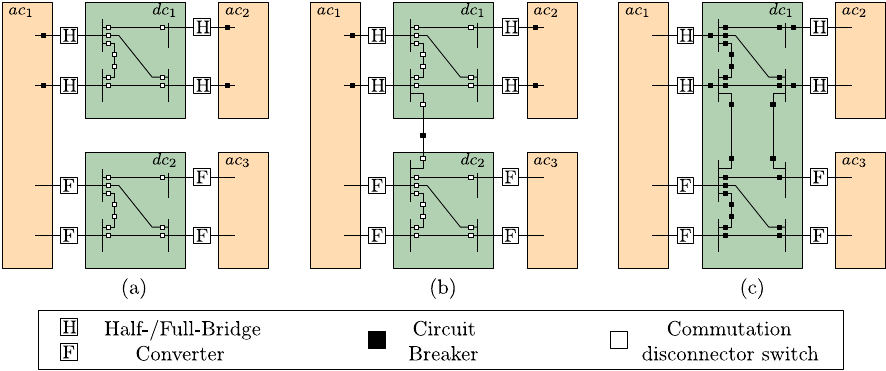}
    \caption{Example fault-clearing strategies in an extended HVdc grid: (a) a nonselective fault clearing using ac circuit breakers (dc1) or fault-blocking converters (dc2), (b) a partially selective fault clearing using HVdc circuit breakers or a dc–dc converter, and (c) a fully selective fault clearing using HVdc circuit breakers~\cite{Protection_strategies}.}
    \label{fig:protection_strategies}
\end{figure}

AC Circuit Breakers (ACCB) are widely used to enhance the power system's redundancy, as they guarantee that a fault can be cleared in each line terminal~\cite{Ss_future_hvdc_grids}. As such, in AC systems, each network component is assigned to protection zones that selectively protect the system in case of contingencies. DC Circuit Breakers (DCCB), on the other hand, are considerably more complex and costly~\cite{DCCBs}. As a consequence of this, mainly three protection strategies exist, illustrated in Figure~\ref{fig:protection_strategies}.

\begin{itemize}
    \item{\textbf{Nonselective strategy}: the entire grid is a protection zone. If a component fails in a DC grid, the entire grid, or the whole faulty pole if the grid only contains full bipoles, is brought to zero power and re-energized after the faulted component has been isolated. Common DC protection components used for this strategy are mechanical switches, such as disconnector and fast disconnector switches~\cite{Ss_future_hvdc_grids}. This strategy is commonly used by point-to-point HVDC links characterized by a limited amount of power. For DC grid power flows of several GW, a nonselective strategy might imply power imbalances higher than the e.g. European dimensioning incident of 3 GW in case of a fault. Having to almost instantaneously ramp down and up in power with a considerable power capacity would hinder the stability of the whole European grid.  As such, for HVDC grids with injection points into the same synchronous area totaling more than 3~GW, a non-selective protection strategy is not feasible in the European context also considering that the defined maximum loss of infeed is 1.8~GW for Great Britain~\cite{SSQS}, and 1.4~GW~\cite{Nordics} for the Nordic area, respectively.}
    \item{\textbf{Fully selective strategy}: the DC grid is protected as the AC one, with DC Circuit Breakers (DCCB) protecting every HVDC link. Since DCCBs are still expensive and considerably larger than ACCBs, the fully selective strategy is not likely to be used for any EEH project unless strictly needed (and economically feasible).}
    \item{\textbf{Partially selective strategy}: the DC grid is split into subparts which are each protected with a nonselective strategy. As a result, the impact of faults on the DC grid is limited while keeping the investment costs for DC protections substantially lower than for the fully selective strategy. Especially two types of DC protection components are required for this strategy, namely i) mechanical switches without fault-current interruption capability within the protection zones and ii) equipment with fault-current interruption capability in the interconnections between the protection zones~\cite{Ss_future_hvdc_grids}.}
\end{itemize}

\subsubsection{Busbar arrangements}
\begin{figure}
    \centering
    \includegraphics[width=0.8\linewidth]{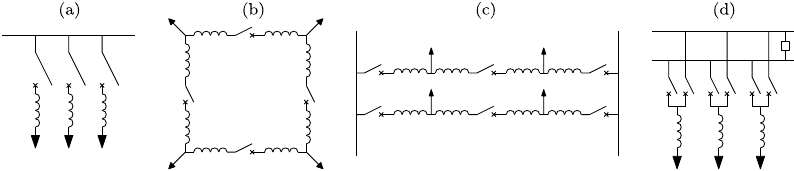}
    \caption{An example of busbar topologies and arrangements for breakers and current-limiting inductors: (a) a single busbar/single breaker, (b) a ring bus, (c) a breaker-and-a-half scheme and a (d) double busbar~\cite{Ss_future_hvdc_grids}.}
    \label{fig:busbar_arrangements}
\end{figure}

Different protection strategies imply that the DC (and AC) busbars need to be adapted to host the protection equipment and/or be able to reconfigure their topology according to the EEH's design choice and goal. Common AC substation layouts such as the single busbar/single breaker, ring bus, breaker and a half or double busbar configurations shown in Figure~\ref{fig:busbar_arrangements} can be extended to DC grids depending on the need~\cite{Ss_future_hvdc_grids}. As mentioned earlier, the increased selectivity and redundancy eventually lead to higher investment costs. In this sense, it is important to select the most efficient protection strategy starting from the early stages of the planning process and include the costs and benefits related to the different protection strategies in the whole planning and design optimization process. Therefore, possible extensions should be foreseen in the EEH design since the beginning of the project, as the space and insulation distances needed by DC equipment are considerably higher and more expensive than for AC equipment. 

\begin{tcolorbox}
[width=\linewidth, sharp corners=all, colback= white!90!black]
\textbf{Constraint}: The co-optimization of the network layout and protection strategy must consider operational security constraints, such as the maximum loss of infeed.\\
\\
\textbf{Consideration}: The cable configuration, MMC converter type and protection strategy choice influence the maximum available power transfer of a multi-terminal DC grid. The EEH's flexibility, redundancy, and extendability should be prioritized, but case-specific arrangements depend on the project's general goal. 
\end{tcolorbox}

\subsection{Costs} \label{sec:costs}
Given the multi-GW size and the amount of network components involved in EEHs, these projects are extremely capital-intensive. For this reason, a careful assessment of the total costs and market conditions related to them is necessary to avoid sub-optimal investments. 

This section elaborates on the Capital Expenditures (CAPEX), Operational Expenditures (OPEX), space requirements, and losses requirements in EEHs. Given the current volatility of the market and uncertainty related to the technology readiness level and large-scale production of some key components, reliable cost figures are difficult to derive~\cite{IEA_2025_constraints}. For this reason, we refer to relevant reports that provide indicative cost figures throughout this section. However, it should be noted that both costs, and particularly market prices, are highly volatile at the time of writing this paper, and might have local differences (labour costs, accessibility to critical resources, carbon cost, etc.) in different areas of the world, but the general reasoning behind the Section can be applied to every EEH project.

\subsubsection{CAPEX analysis and market fitness} \label{sec:capex}
CAPEX are the upfront costs paid by an investor to build the structure and electrical equipment of an EEH, including financial interests to be paid to the actors lending the capital. The “IEEE Technical report PES-TR86 - Studies for
Planning HVDC” ~\cite{Stu_for_plan_HVDC} is used as the main reference to list the main factors related to the CAPEX of an EEH.   
\begin{itemize}
 \item{\textbf{Development: Planning, Consent, Front End Engineering and Design (FEED)}: FEED analyses are essential to assess a project's overall engineering-related costs. While the general best practices for budgeting large-scale (or mega) projects are extensively discussed in the literature~\cite{Flyvberg_1,Flyvberg_2}, some peculiar aspects related to EEHs can be listed, especially when related to offshore investments. CAPEX costs for EEHs vary substantially among different installations depending on several project-specific factors, such as regulatory framework, variability in cost components, TSOs and regulators' involvement, optimal technical decisions, location of the installation, etc. Additionally, separating the converter supply and civil works into different contracts increases the overall costs (if there is no abuse of market power) since both parties require manpower for activities such as project management and site supervision. When the effort needed for coordination, management, and control interfaces increases, the enhanced complexity of the project leads to more risk exposure, i.e., more costs. In addition, the lack of public data about general offshore installations increases the uncertainty around EEH and makes the benchmarking of the costs with similar projects impossible. A general rule of thumb for TSOs is that the cost of physical areas/volumes (especially offshore) is the main cost component for FEED analyses. Therefore, the space occupied by equipment and the project's general offshore footprint needs to be minimized.}
 \item{\textbf{Procurement of equipment and components, market fitness}: 
 Given the complexity of the technology, high investment costs, and highly skilled resources needed, only a handful of manufacturers can currently build HVDC cables and (VSC-MMC) converters at the moment~\cite{IEA_2025_constraints}. In addition, no new entrant is expected to appear in the market soon, as the industry's investment and operational costs are extremely high. Furthermore, entering the market requires highly skilled personnel, which is currently difficult to acquire. Moreover, the low offer and increasing demand for HVDC links around the globe, cause high market volatility and a push towards standardized solutions for the converter manufacturers to speed up their production and create a supply chain around them, avoiding custom-built solutions. As a result, even if the HVDC configuration and protections used in an EEH should be optimized according to the general goal of the project, there is a push towards standardization of a common solution, such as 2GW 525~kV (higher voltages are possible, such as 640~kV~\cite{CIGRE_NKT} and 800~kV~\cite{800_kV}) HVDC bipoles with DMR~\cite{Tennet_2GW}, which can lead to a lower price for this configuration even if they involve more components and conductors.
}
\item{\textbf{Logistics and installation}:
The installation costs of EEHs are directly related to the location of the project and the existing infrastructure around it. Parameters such as the types of soil, recurrence of heavy storms, presence of wildlife in the area, harsh weather conditions, limited physical accessibility, etc., increase the project's total costs. While a general rule of thumb is that the more remote an installation is from the load centers, the higher the installation costs would be, offshore investments tend to be the most expensive, especially when in deep waters with limited accessibility due to constant strong winds around them. Onshore investments have lower installation costs, but their public acceptance is extremely low, as the population dislikes overhead lines (more) and cables (less) passing close to their properties~\cite{NEUKIRCH_Grid_acceptance}. In this sense, the practical installation of HVDC cables and possible obstacles on the designed route are influencing the overall costs. Moreover, particular attention should be paid to laying the cables in areas with rocks, volcanic phenomena, etc.~\cite{cables_rocks}, as damaging the cable leads to substantial delays in the commissioning of the projects.

Onshore projects are logistically more easily accessible than offshore ones, as the latter requires several boats to carry the equipment and build the EEH offshore. In the case of offshore wind, building the construction facilities on the coast next to a harbor assures that the wind turbines and electrical equipment are delivered to the EEH location more easily. Nevertheless, the weather conditions of the area need to be studied to find possible windows in which the equipment can be installed safely.}
 \item{\textbf{Interest During Construction}: Typically, interest rates are related to the inherent risk of the project. A reliable timing of the project is necessary to avoid delays, which lead to considerably higher interest at the end of the project, i.e. additional investment costs.}
\end{itemize}

Note that recent European system studies, such as the Elia blueprint for 2050~\cite{blueprint}, include in their scenarios HVDC investment overnight costs in the range of 1680–3800 €$_{2022}$/MW/km for onshore links and of 1680–3000 €$_{2022}$/MW/km for offshore links. Regarding the AC/DC converter station costs, the onshore costs range is 250–325~M€$_{2022}$/GW, while the offshore one is 550-700~M€$_{2022}$/GW, including the offshore platform required to hold the converter. The cost of other (expectedly expensive) components, such as DCCBs, remains highly uncertain, as these technologies are not yet deployed in the European system. 
It should be noted, however, that these estimates are subject to significant uncertainty due to the current market volatility, and future cost levels may deviate substantially from the reported ranges. With the expected future improvements in EEH construction and delivery, and EEH installations in several areas in the world, the cost figures for each component will likely be more mature. For this reason, local system operators should thoroughly analyze the cost figures in their region to make both an economically efficient and technically sound choice.

\begin{tcolorbox}
[width=\linewidth, sharp corners=all, colback= white!90!black]
\textbf{Consideration}: An EEH should meet its power aggregation and transmission goals at the lowest possible cost while meeting the existing technical and space requirements. 
\end{tcolorbox}
       
\subsubsection{OPEX analysis}\label{sec:opex}
Operational expenses (OPEX) are costs related to the daily operations of an EEH and, in general, to all the operational expenses during its lifetime that were not included in the CAPEX assessment. Two OPEX-related categories from~\cite{Stu_for_plan_HVDC} can be applied to the context of EEHs:
\begin{itemize}
    \item{\textbf{Designed lifetime}: given the significant investment costs, EEHs are expected to operate for several decades. For example, the Princess Elisabeth Belgian energy island~\cite{Eliaenergyisland}'s structure is built with a 100-year time horizon. As the electrical installations have a considerably lower lifetime, the substations and converter stations are supposed to last for at least 40 years~\cite{Eirgrid}. The auxiliary low-voltage equipment has instead lower requirements, such as 16-20 years, being easier to replace and less expensive.
    Additional reinforcements to the EEH's structure should be foreseen for critical elements located in extreme conditions and difficult to replace, such as high-voltage offshore transformers. In such cases, a trade-off needs to be made between a more robust and thus expensive design (CAPEX) and more recurrent maintenance (OPEX).}
    \item{\textbf{Cost of unplanned and planned maintenance actions}: every unplanned maintenance action results in an economic loss due to the reduction of power transfer through the EEH, as the installations are unmanned and need to be available 24/7. The repair time of such faults might be increased by harsh weather conditions, especially for offshore projects, due to the difficulty of reaching the installation in adverse weather conditions.  Spare components of key elements should therefore be foreseen on the EEH. Similarly, maintenance periods need to be optimally planned to minimize the out-of-service hours.  In this sense, choosing highly reliable systems, i.e., with low failure rates, reduces the mean time to repair and increases the availability of the overall project, as discussed later in the paper in Section~\ref{sec:RAM}.}
    \item{\textbf{Cost of Frequency Restoration Reserve increase}: even if the most impactful incident of an EEH can be lower or equal to the dimensioning incident (FCR of the connected grids), its impact can be higher than the LOI or loss of load (LOL) resulting from any other incidents within a control area. As such, the integration of an EEH into a control area can lead to the system operator having to purchase more mFRR and/or aFRR services. The cost difference between the frequency restoration reserves that have to be purchased by a system operator with and without considering the EEH can be attributed to the EEH as OPEX costs. Depending on the cost to increase these reserves, it could be better to limit the capacity of the EEH.}
\end{itemize}

\begin{tcolorbox}
[width=\linewidth, sharp corners=all, colback= white!90!black]
\textbf{Consideration}: The designed lifetime and planned maintenance actions are design parameters depending on the characteristics of the EEH. The (potential) unavailability of the project should be minimized by organizing effective planned maintenance actions and investing in reliable components, especially if the EEH is placed in harsh weather conditions.

The capacity of the EEH can lead to increases in the required frequency restoration reserves. These need to be considered in the OPEX estimates of a candidate project and might influence the EEH’s optimal capacity.
\end{tcolorbox}

\subsubsection{Space requirements}\label{sec:costs_space}
As previously mentioned in the text, the cost of the equipment offshore, or offshore cost, is the main component of the investment costs for offshore installations all over the world. First, when considering building new EEHs, three main categories of projects can be identified from the space requirements perspective. According to each category, different design constraints are applied to a new EEH project.
\begin{itemize}
    \item{\textbf{Construction of an EEH with limited surface area}: the total surface area of the EEH is fixed because of space limitations given by, e.g., existing infrastructure, nature reserves, or political borders. The EEH design choices must satisfy the requirements for generation and transmission capacity without exceeding the space limits. 
}
    \item{\textbf{Construction of an EEH with limited installed generation and/or transmission capacity}: the capacity of the EEH is constrained because of limitations in the total generation and/or transmission capacity that can be installed. The focus is on achieving the grid functions expected by the project. Nevertheless, the total surface area, i.e., the investment costs, should always be minimized. 
}
    \item{\textbf{Construction of an EEH with no limited surface area and capacity constraints}: the EEH's design can be optimized without pre-existing limitations regarding its area or installed capacity. The EEH's design is optimized according to the project's overarching goal discussed during its planning stages.
} 
\end{itemize}
 
\begin{figure}
    \centering
    \includegraphics[width=0.45\linewidth]{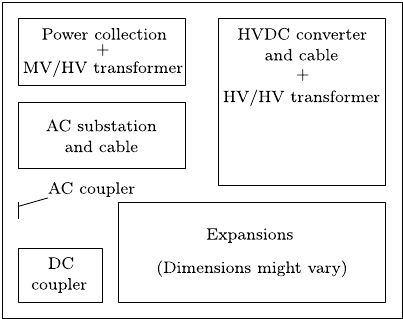}
    \caption{Main building blocks of an electrical energy hub. The ratio between the areas of the components are based on realistic data from~\cite{DTU}.}
    \label{fig:space_requirements}
\end{figure}

Based on the Belgian~\cite{ELIA_PEI} and Danish~\cite{Bornholm,DKislands} proposed EEHs projects, six main building blocks and their components are distinguished. Note that the dimensions of the elements are taken from publicly available sources~\cite{DTU} and are related to components that have yet to reach maturity in most of the cases. While, the dimensions of each network component can be different depending on its manufacturer, an estimate of the dimension ratios between the several components is displayed in Fig.~\ref{fig:space_requirements}. 
\begin{itemize}
    \item{\textbf{Wind collection}: collection point of the AC cables transferring power from the Offshore Wind Farm (OWF) to the EEH and AC infrastructure to raise the AC Medium Voltage (MV) from the OWF (32 kV, 66 kV or possibly 132 kV in the future) to High Voltage (HV) through a transformer. Additional elements that can be present in this building block are AC Pre-Insertion Resistor (AC-PIR), AC harmonic filters, and AC switchgear. According to~\cite{DTU}, the AC substation for the collection of the wind farm has a dimension of roughly 580 m\texorpdfstring{$^{2}$}{}, while the MV/HV transformer 128 m\texorpdfstring{$^{2}$}{} is 4.5 times smaller.}
    \item{\textbf{HVDC converter and cable}: VSC-MMC 525 kV converters are proposed to be the standard for offshore HVDC cables. These converters need the AC voltage to be in the range of 250-300~kV. An HV/HV transformer is therefore required to increase the AC voltage to the desired level. Additionally, the converter might require DC-PIR~\cite{DC_PIR}, DC braking system~\cite{DC_BS}, DC reactor~\cite{DC_reactor}, and DC cable discharge~\cite{DC_discharge} devices depending on the project. The HVDC cables for offshore transmission in the future, based on TenneT's 2~GW plan~\cite{Tennet_2GW}, are expected to have a DMR. Furthermore, the converters are the largest electrical component. Depending on the manufacturer, their valve halls have dimensions of roughly 32~m~x~59~m~x~18~m for the 500~kV MMC base case and 25~m~x~45~x~12~m for the 100 kV MMC with parallel arms and MMCs in parallel~\cite{MMC_dimensions}. According to~\cite{DTU}, the whole converter occupies an area of roughly 2250 m\texorpdfstring{$^{2}$}{}, with the HV/HV transformer being 128 m\texorpdfstring{$^{2}$}{}.}
    \item{\textbf{AC substation and cable}: for installations close to shore such as the Princess Elisabeth energy island~\cite{ELIA_PEI}, AC cables can still be used to transfer the power onshore. Before the AC cable, the voltage from the OWF collection points needs to be increased to 220 kV with an MV/HV transformer which is estimated to have the same length and width as the HV/HV transformer~\cite{DTU}. The AC substation, mentioned previously in the section, has an estimated area of 580 m\texorpdfstring{$^{2}$}{}. As a comparison, existing AC 220~kV platforms in the Baltic and North Seas have a surface of 51~m~x~31~m (1581~m$^{2}$) and a height of 35~m for the Baltic Eagle 476~MW offshore wind farm~\cite{Baltic_eagle}, and a surface of 58~m~x~32~m (1856~m$^{2}$) and a height of 26~m for the 700~MW Borssele offshore wind farm~\cite{Borssele}. Examples of equipment in the AC substation are AC switchgear, AC harmonic filters, AC protections, and AC-PIR.}
    \item{\textbf{AC coupler}: AC coupling equipment is used to connect and disconnect part of the EEH via AC switchgear, similarly to what is shown on the left-hand side of Figure~\ref{fig:AC_DC_coupling}.
    \begin{figure}
        \centering
        \includegraphics[width=0.55\linewidth]{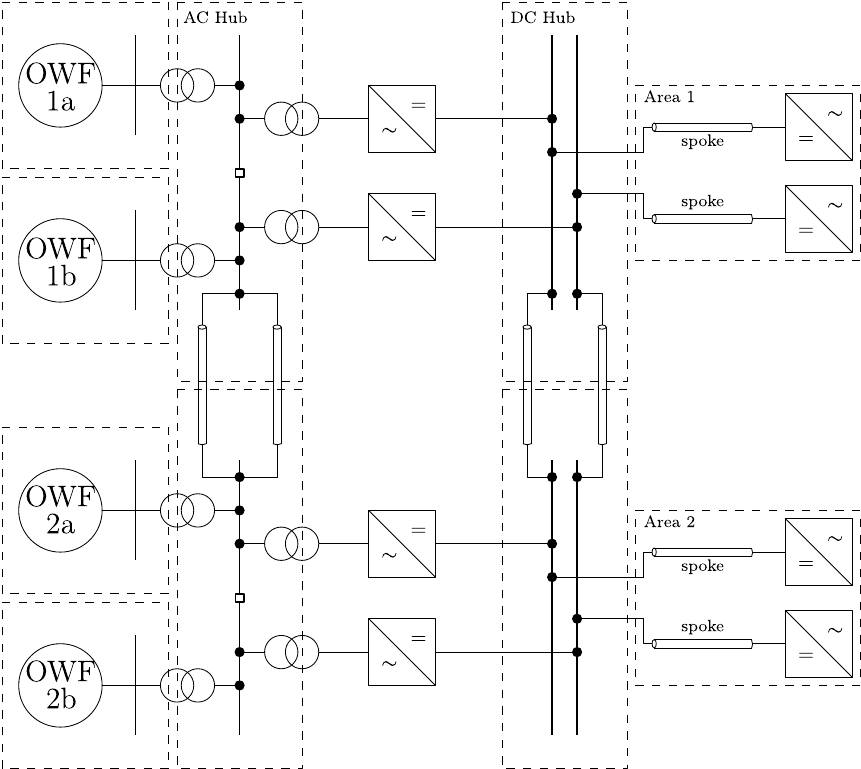}
        \caption{Hybrid hub solution from~\cite{DTU} in which AC (left) or DC (right) coupling is possible.}
        \label{fig:AC_DC_coupling}
    \end{figure}
    While the AC coupling is useful for rearranging the AC busbar topologies of an EEH, the space required for these installations is trivial compared to other parts of the project.}
    \item{\textbf{DC coupler}: DC switching units are used to DC-couple different HVDC cables in an EEH, as shown on the right-hand side of Figure~\ref{fig:AC_DC_coupling}. The network components being used are the same as the ones mentioned in the ``expansions'' part of the section but, in this case, only the DC switchgear is considered. The same reflections about the DC protection strategies and space needed for the equipment are valid for this point.}
    \item{\textbf{Expansions}:
    Eventual expansions towards other EEHs control zones need to be considered in the EEH's original plan. For example, the first multi-terminal VSC HVDC grid in Europe, the Caithness Moray Shetland project, was designed to allow further expansion in the point of coupling of its terminals~\cite{CMS}. Depending on the chosen DC protection strategy and the EEH's operations in conjunction with the overall system, possible DC protections include DCCBs or DC disconnectors. 
    
    On the one hand, with a DCCB, a DC fault can be interrupted at any point in time, and the faulted line is isolated without power loss in the other DC lines~\cite{Mudar_PhD}. 
    
    On the other hand, using DC disconnectors implies that the power through the entire DC grid needs to be brought to zero in case of a DC fault, with a nonselective protection strategy. Afterwards, the faulted line is removed, and power is transferred again through the DC grid. Since DCCBs are expensive and large network components, the DC-preventive decoupling strategy~\cite{Dullmann_2023,Dullmann_2024} has been proposed to avoid their use in small DC grids, and it has been techno-economically evaluated against DCCBs in~\cite{Van_deyck}. Following this strategy, a given DC system is decoupled without loss of power into several zones when the total power flowing through the DC grid is above the dimensioning incident discussed in Section~\ref{ch:NOS}. Nevertheless, the maximum amount of power that can be transferred during the DC grid is reduced.
    
    The operator of the EEH must therefore choose between a more robust DCCB or a preventive DC-side decoupling strategy with DC disconnectors, which results in less power being transferred but smaller and less expensive equipment, as~\cite{DTU} reports the DCCB to occupy about 400~ m\texorpdfstring{$^{2}$}{}, while DC disconnectors are significantly smaller. Note that for more extended Multi-Terminal DC (MTDC) grids, the DCCB is arguably the preferred choice to guarantee the secure operations of the power system.}
\end{itemize}

Note that auxiliary equipment such as the harbor for service boats, meeting rooms, safety protection walls, safety distance from the walls, etc. are not discussed in this section, but they require a minor percentage of the area of the EEH as well. While the dimensions of the various EEH building blocks in the Section are drawn by a Danish report from DTU~\cite{DTU}, they should be comparable to the ones for EEHs in different parts of the world. Depending on the location of the EEH, fewer or more civil constructions are needed, e.g., for an installation in a shallow sea (bottom-fixed) or in deep waters (floating platform).   

\begin{tcolorbox}
[width=\linewidth, sharp corners=all, colback= white!90!black]
\textbf{Consideration:} The maximum dimensions of the different building blocks of an EEH are used to select the best combination of network components for a given space or power capacity.
\end{tcolorbox}

\subsubsection{Losses}\label{sec:losses}
Electrical losses are a typical design requirement from TSOs to the components' manufacturers. Following the Joule losses' law \texorpdfstring{$P_{l} = R \cdot I^{2}$}{}, there has been a tendency to reach higher voltage levels and lower currents in the last decades. The main technical constraint related to losses is the safe evacuation of the heat coming from the key electrical components. As methods to compute the losses for e.g. cables and MMC converters are well known~\cite{MMC_losses}, each component should be designed to operate safely in a given temperature range for most of its lifetime, being able to control the amount of losses while the cooling equipment prevents it from overheating. From the techno-economic perspective, power losses can be a significant cost factor as the losses are accumulated over the lifetime of the assets of several decades. In this respect, the costs associated with power losses for HVDC systems can be in the order of 10~\% of the total expected CAPEX accumulated over the lifetime of the asset, as computed in~\cite{hakan_powertech} for a 300 MW offshore wind connected via VSC HVDC to the onshore system. 

Therefore, it is beneficial to investigate trade-offs between the cost of measures for reducing system losses and the anticipated costs for losses themselves. As an example, using submarine cables with higher cross sections inherently will lead to higher CAPEX of equipment due to the increased use of copper or aluminum, as well as higher installation costs due to the cable weight. However, such costs might be justified if the lifetime loss reduction due to the higher cross-section is economically acceptable. 

\begin{tcolorbox}
[width=\linewidth, sharp corners=all, colback= white!90!black]
\textbf{Consideration:} The losses in the system are a design choice for system operators. They can account for some percentage of the total investment-related costs over the project's lifetime. 
\end{tcolorbox}

\subsection{Electricity market design} \label{sec:el_market}
While discussing different market mechanisms is beyond the scope of this work,  electricity market designs are briefly discussed in this Section, as they turn out to be one of the most relevant parameters for the effective implementation of grid expansion planning projects~\cite{Hardy}. In fact, the choice of the market design in which the EEHs are operated has a large influence on the benefits to the socioeconomic welfare brought by the EEH. 
When proposing a market design, a fundamental trade-off exists between \textit{market efficiency}, \textit{incentive compatibility}, \textit{cost recovery}, and \textit{revenue adequacy}, as no market-clearing mechanism can simultaneously satisfy all four properties~\cite{Hurwicz,Jalal_markets,Mitridati,Myerson}. The market design must therefore explicitly prioritize among these properties according to the overarching goal of the EEH project.

In the context of EEHs, the overarching goal is to minimize the cost for society while guaranteeing that the owners of the generation assets are remunerated for their investments.  Therefore, the market design decision should consider the interests of as many stakeholders as possible to select the most efficient design according to the final goal of the EEH project. Nevertheless, the market design is closely tied to the regulatory framework of the country/area where the EEH is built. Therefore, the market design shapes the operating regime of an EEH (and thus influences design choices like capacity sizing and interconnection topology) without directly bounding any single design parameter. In this sense, the next paragraph brings the example of the current ongoing discussions about the cost-remuneration mechanisms and market designs of EEH projects in the North Sea, but this should be considered as an explanatory example rather than a technical constraint, as the remuneration mechanisms for investors in EEHs need to fall within the legislation of the area where they are built.

Currently, EEHs and HVDC links in general are extremely capital-intensive projects, but will likely benefit more than one system operator, especially in locations like the European North Sea. Based on this fact, and given the fragmented nature of the European power system from a system operator perspective, the first regulations for cost- and benefit-sharing have been proposed by the European Commission~\cite{CBCA_1,CBCA_2}. Specifically in the North Sea, the Offshore TSO collaboration~\cite{OTC_II} investigates the possibility of a ``Joint Regional Cost Sharing''~\cite{OTC_IV}, intending to involve the parties sharing the costs in the offshore grid planning and cost-sharing stage, and to provide a transparent cost-sharing decision providing the right remuneration for all the involved parties, also through the use of metrics, e.g. ex-ante metrics (prior to any final investment decision), ex-post (after a final investment decision) metrics or a combination of the two. While the interested reader is referred to~\cite{OTC_IV} and ongoing work from the Offshore TSO Collaboration for a comprehensive description of methodologies to allocate costs and benefits among different stakeholders, it is worth emphasizing that such initiatives are instrumental in refining existing approaches to identifying system needs~\cite{ENTSO_E_system_needs}. As such, they are expected to play a key role in determining the future expansion of the European power system.

\begin{tcolorbox}
[width=\linewidth, sharp corners=all, colback= white!90!black]

\textbf{Consideration:} The market design governing an EEH determines which interconnection topologies and ownership arrangements are financially viable, and therefore constrains the set of technically feasible configurations that can be considered. In particular, the chosen cost-remuneration mechanism (e.g., ex-ante or ex-post cost sharing) affects the optimal capacity and number of interconnections of the EEH, and must be aligned with the regulatory framework of each connected control area before a final investment decision can be made.
\end{tcolorbox}

\subsection{Future proofness \& modular expandability} \label{sec:future}
\subsubsection{Expandability}
Expandability aims to ensure that both predictable and unpredictable future developments can be integrated into a system with minimal adaptations to the already existing infrastructure. While expandability should be enabled at marginal additional costs, minimizing prior investments benefits the economic feasibility of an EEH. 
Therefore, from a system perspective, developers need to find the optimal design to avoid sunk costs while making sure ``no regret'' investment decisions are identified and executed on time.

\subsubsection{Achieving modular expandability}
As introduced in Section~\ref{sec:capex}, the standardization of HVDC technologies helps manufacturers speed up their production and create a supply chain around their key network components. As a result of this standardization, modular EEHs can be established following the building blocks listed in Section~\ref{sec:costs_space}. The benefits brought by modular expansions~\cite{Modularity} in EEHs are:
\begin{itemize}
    \item Economy of scale and optimized supply chain, resulting in faster construction and delivery times~\cite{WEF}.
    \item Enhanced flexibility of the whole project.
    \item Avoidance of sunk costs due to i) production inefficiencies, ii) unknowns due to limited knowledge of sensible information, and iii) non-efficient planning for ad-hoc solutions.
    \item Steady incremental growth.
\end{itemize}
Among the benefits, the avoidance of sunk costs is often overlooked. While production inefficiencies and non-efficient planning are general issues, the effective expansion of planned MTDC projects depends on the ability of components supplied by multiple vendors to be compatible and work together under varying operational modes, defined as ``interoperability"~\cite{InterOpera.D4, wang2021multi}.
Without agreements on key interoperability topics the design process is challenging due to both technical reasons, e.g. difficulties in predicting and ensuring the stable integration of components without experiencing interactions or incompatibilities, as well as organizational. The reader is referred to~\cite{Definitions_paper} for a thorough explanation of the interoperability issue in EEHs.

Moreover, EEHs' expandability depends mainly on the location where the project is built. In the North Sea case, the political agreements between the North Sea countries, the high capacity factors from offshore wind, and the lack of pre-existing infrastructure between the countries make it the ideal case to foresee further connections to other EEHs. If an EEH is in a smaller and more isolated area far from other control areas, the interconnection will most likely be a point-to-point HVDC link. Therefore, we identify the following drivers for EEH's expandability:
\begin{itemize}
    \item Presence of existing infrastructure in the same area as the EEH.
    \item Physical location
    \item Agreements with other control areas
    \item RES capacity factors in the area
\end{itemize} 

Optimization models can be developed to ensure that the EEH's building blocks occupy the lowest area possible while achieving the project’s goal. The co-design of planning, operations, control, and protection of MTDC grids is advised to maximize the integration of RES in the existing power system. When future extensions are considered, feasibility studies such as optimal power flow simulations, electromagnetic transient studies, and fault analyses should be performed to guarantee the safe and reliable operations of the EEH and the already existing grid as a whole. The output of these studies might result in requirements for some network components such as, e.g., converter fault-ride through and multi-vendor interoperability.

\begin{tcolorbox}
    [width=\linewidth, sharp corners=all, colback= white!90!black]
        \textbf{Consideration:} Expandability is one of EEHs' key purposes for the power grid of the future. It consists of the integration of adequate provisions for interface facilitation in the system design. A balance between the controlled costs of prior investments and those of later uncertain developments needs to be found to achieve economic efficiency.
\end{tcolorbox}

\subsection{Reliability-Availability-Maintanability} \label{sec:RAM}
As all the network elements in a power system, EEHs are subject to failures and other unpredictable events that have the potential to make them partly or fully unavailable for a certain period of time, dependent on the type of fault. On the one hand, Reliability-Availability-Maintainability (RAM) compute an inherent trade-off between the simplicity of a system that needs to remain as lean as possible to limit the probability of occurrence of undesirable events. On the other hand, the application of the right level of subdivision and integration of redundancy limits the impact of such fault events.
As such, RAM is an overall system optimization problem that balances the three main features:

\begin{itemize}
    \item \textbf{Reliability}, defined as \textit{``the probability of a device or system performing its purpose adequately for the time intended
under the operating conditions encountered''}~\cite{4074107}. Reliability is linked to two main aspects, adequacy and security.
    \begin{itemize}
        \item Adequacy, or static reliability, is the ability to ensure that there is enough generation and transmission capacity to meet the demand in a given power system.
        \item Security, or dynamic reliability, is the ability to ensure that a given power system is resilient against abrupt disturbances, such as faults, short circuits or loss of components. 
    \end{itemize}
    Ideally, the simpler and better protected the system is, the lower the failure rate. 
    Common indicator values for different EEH building blocks are, e.g., forced outage rates of about 3–4 per year for a single pole and 0.1 per year for a bipole in LCC HVDC converter stations according to the CIGRE Technical Brochure 713~\cite{CIGRE_TB713}. Regarding submarine cables, the average annual failure rate is indicated as 0.096 for 100~km, with approximately 60 days of repair time~\cite{CIGRE_TB713}. In addition, the annual failure of a submarine cable is 0.07 failures per 100km~\cite{Rel_1991}. The reader is referred to~\cite{HEYLEN2} for a complete list of reliability indicators, such as the System average interruption frequency index (SAIFI), the Average system interruption frequency index (ASIFI), etc..
    \item \textbf{Availability}, measuring the probability of a given network component to function in a given period. As a rule of thumb, the lower is the curtailment of the rated capacity, the higher the system availability. For example, LCC HVDC converter stations are indicated to have an energy availability of about 98–99\% in the same CIGRE Technical Brochure 713~\cite{CIGRE_TB713}. The same brochure indicates an energy availability of approximately 96.5\% for radial VSC-HVDC, with the inactive time mainly due to forced outages. For offshore installations subject to harsh weather conditions and limited accessibility, availability figures toward the lower end of this range are more representative, underscoring the importance of redundancy provisions and optimised maintenance scheduling in the EEH design. These availability targets have a direct bearing on the OPEX estimates discussed in Section~\ref{sec:opex}, since a reduction in energy availability translates directly into lost transmission revenue and potentially increased frequency restoration reserve costs for the connected control areas. 
    \item \textbf{Maintainability} refers to the \textit{``Measures taken during the development, design, and installation of a manufactured product that reduce required maintenance, manhours, tools, logistic cost, skill levels, and facilities, and ensure that the product meets the requirements for its intended use \cite{Maintain}''}. Indicatively, the higher the subdivision, redundancy, and accessibility of the system, the higher its maintainability. For offshore installations, CIGRE TB 379~\cite{CIGRE_TB379} highlights that the mean time to repair for submarine cable faults is strongly governed by vessel availability and weather windows rather than the fault itself, with offshore repair campaigns typically lasting between two and eight weeks. This has a direct implication for EEH design: critical components that are difficult or slow to repair offshore, such as HVDC cables and offshore transformers, should either be specified with higher reliability targets or be provided with sufficient redundancy so that the system can continue to operate at reduced capacity during the repair period.
\end{itemize}

\begin{tcolorbox}
[width=\linewidth, sharp corners=all, colback= white!90!black]
\textbf{Consideration:} The availability of EEHs heavily influences their benefits to the socioeconomic welfare and thus needs to be maximized within the boundaries of reasonable capital expenditures. It is a natural balance between system simplicity and integrated redundancy.
\end{tcolorbox}

\subsection{Sustainability} \label{sec:sustainability}
EEHs are by definition key enablers of asset rationalization and lead to the reduction of $CO_{2}$ emissions through the transmission of sustainable electricity. Nevertheless, the impact of such major infrastructure on the environment is substantial. They thus require additional efforts to mitigate negative impacts and provide support and preservation to nature for a sustainable environment.

Sustainability is promoted through diverse laws and regulations to which asset owners need to comply. As such, permitting procedures integrate strong requirements related to environmental impact assessment~\cite{BGA}. In addition, direct financial incentives also further reinforce efforts towards carbon neutrality. Sustainability measures can be clustered into 3 main domains of action: climate, biodiversity, and societal:
\begin{itemize}
    \item \textbf{Climate}: $CO_{2}$ emissions are present from the planning to the decommissioning of an EEH project. These emissions need to be tracked transparently, reduced to the maximum extent, and compensated. The most impactful and traceable activities concern equipment production, transport, and installation.
    
    First, the amounts of material and, more specifically, of metals are to be minimized. As such, preference goes towards technologies offering the highest electricity transport capacity per unit of total equipment weight. For HVDC systems of a specific rating, different technologies may exist and their climate impact assessment will likely require a trade-off between the voltage level and the current for:

\begin{itemize}
\item The converter stations, in which structural quantities play a major role and directly depend on insulation distances, i.e. voltage levels.
\item The HVDC cable system, in which metal conductor quantities can amount to very large figures, especially for very long interconnectors, depends on its current rating.
\end{itemize}

Second, the energy consumption and emissions types during the construction of EEHs are related to the amount of necessary work and the means to perform these operations. Ensuring minimization of transport and installation works, together with the use of adapted machinery whose own climate impact throughout its life cycle is the lowest available on the market, are relevant indicators of a positive climate approach. 

Third, the use of gas-insulated high-voltage equipment to reduce the footprint of converter stations has to be treated cautiously. In the context of the phase-out of the fluorinated greenhouse gases~\cite{SF6_regul,SF6_regul_II}, the use of SF$_{6}$ as the most mature gas insulating technology currently continues to represent a risk of large additional emissions to be balanced with its associated space gains and positive impact on the number of structural materials. More concretely, from 1 January 2028, Regulation (EU) 2024/573~\cite{SF6_regul_II} prohibits the putting into operation of new or extended SF$_{6}$-containing ``\textit{high voltage electrical switchgear from 52~kV up to and including 145~kV, and up to and including 50~kA short circuit current}''. The same regulation includes, from 1 January 2032, ``\textit{high voltage electrical switchgear of more than 145 kV or more than 50 kA short circuit current, with a global warming potential of 1 or more}''. This regulation makes the early selection of non-fluorinated alternatives a near-term planning constraint.

Lastly,  the overall power losses (mostly by heat dissipation in electrical resistances) can add up to significant figures in the overall lifetime of an EEH project. Therefore, they must not be neglected when assessing the overall climate impacts of the proposed solutions, as mentioned in previous Section~\ref{sec:losses}.

\item \textbf{Nature and Biodiversity}: in addition to the global climate impacts, sustainable projects should consider their influence on direct natural surroundings. Existing and future fauna and flora can in fact be considerably impacted by the new infrastructure and related construction works. In practice, a set of preliminary studies focused on nature inclusion and promotion of biodiversity should be performed to identify the necessary measures early on and plan for these to reach a nature-inclusive design boosting the ecological and environmental value from the start of the project~\cite{Elia_nature}. Among others, the following levers represent primary focus aspects that can have a considerable influence on the wildlife in the EEH region:
\begin{itemize}
\item \textbf{Cable routing}: the selection of appropriate cable routes, bundling strategy, trench design and burial depth.
\item \textbf{Offshore structural elements}: the selection of appropriate foundations location, materials and design characteristics.
\item \textbf{The use of hazardous products}: the limitation and mitigation of leaks and spills of harmful products such as oils, heavy metals or other chemicals.
\end{itemize}

Even though a nature-inclusive design and project planning are fundamental, specific monitoring throughout the entire process is necessary, as well as a proper decommissioning strategy at the end of the asset's life cycle. This ensures that the parts of the EEH are treated following the circular economy's rules and best practices~\cite{circular}.

\item \textbf{Society}: EEHs are developed in the general interest of society. Their practical implementation should as such also consider the perspective of the exposed public in the direct vicinity of the new assets. In addition, the well-being of the construction and operation workers involved in the various stages of the project, i.e. engineering, manufacturing, transport and installation, operation, etc., should always be considered as a priority. Even though a case-by-case investigation is suggested for the numerous applicable norms, industry standards, and local regulations on the topic, general due diligence is expected on the health and safety of all the workers. By nature, the development and operation of high-voltage electrical infrastructure brings its share of safety hazards that need to be properly mitigated using adequate measures. Among others, audible noise and electromagnetic fields in the low as well as high-frequency ranges are consecutive emissions emanating from (parts of) the electrical infrastructures. Their realistic anticipation during the design stage, detailed monitoring during the production process, and final on-site verification during commissioning and trial operation consist of the main part of the environmental impact assessment of the project.

Moreover, proximity sourcing reduces the amount of transport work with direct climate impact, while promoting the local industry and creating an additional common sense of purpose for the population. 
Therefore, the asset owners need to plan for clear and transparent public communications from the very start of the project to ensure an open dialogue with the community and mutual understanding.
\end{itemize}

\begin{tcolorbox}
[width=\linewidth, sharp corners=all, colback= white!90!black] \textbf{Consideration}: A sustainable EEH requires minimization of its associated $CO_{2}$ emissions, in-depth comprehension and willingness to support the surrounding natural habitat, together with empathy towards the vast community of direct and indirect project participants and stakeholders.
\end{tcolorbox}
\section{Discussion} \label{sec:discussion}
Throughout the previous sections, we discussed several areas of interest to consider when evaluating investments in EEHs, or, more generally, grid expansion planning projects. For each area, we identified the main technical design constraints and considerations to be addressed during the planning of such projects. Table~\ref{tab:constraints_summary} 
includes a summary of all the identified technical design constraints and considerations per area, while classifying each area in three different criticality classes, namely \textit{hard constraints}, \textit{main drivers}, and \textit{key considerations}. Even though all the technical design constraints and considerations need to be fulfilled to advance the planning process of an EEH, \textit{hard constraints} are the most critical for system operators and must be fulfilled before proceeding with the further stages of the planning of an EEH. \textit{Main drivers} are related to the overall operation and control philosophy of an EEH. Different from the \textit{hard constraints}, they are not binding from the system perspective but are key for the efficient and secure use of the EEH. Moreover, \textit{key considerations} are related to the design parameters of different network components. Although arguably less relevant to the overall control of a power system, they can considerably influence the final design of an EEH or its associated investment, and should therefore not be underestimated.

The criterion that assigns the criticality of each constraint/consideration is as follows. An item is a \textit{hard constraint} when it is a binding operational security constraint, a condition on the secure operation of the connected power system that an EEH design must satisfy, or when its selection directly determines the value of such a constraint. An item that is not a binding operational security constraint but governs the overall operation, control, and investment philosophy and the system-level capacity and topology decisions of the EEH is a \textit{main driver}. An item that acts on the design parameters of individual network components, rather than on the operation of the system as a whole, is a \textit{key consideration}. This criterion underlies the ``Constraint/consideration criticality'' column of Table~\ref{tab:constraints_summary}.

\begin{table}[h!]
\fontsize{7.5pt}{9pt}\selectfont
\begin{tabular}{c|c|c|c}
\hline
Identified & Area & Description of the & Constraint/ \\
technical constraint/ & of & technical constraint/ & consideration \\
consideration & interest & consideration & criticality \\
\hline
 \rowcolor{gray_light} & & The grid must always be operated to avoid the presence of any contingency&\\
 \rowcolor{gray_light} \multirow{-2.0}{*}{Network} & Network & with an instantaneous impact larger than the dimensioning incident &\\
  \rowcolor{gray_light} \multirow{-2.0}{*}{operational} & integration  & for the given power system, (e.g., the European maximum FCR level), and with & \multirow{-3}{*}{Hard}\\
    \rowcolor{gray_light} \multirow{-2.0}{*}{security} & & a lasting impact larger than the frequency restoration reserves in any control area. & \multirow{-3}{*}{constraints}\\
\hline
 \rowcolor{gray_light} & & Topological remedial actions and efficient grid expansion have plans & \\
 \rowcolor{gray_light} \multirow{-2}{*}{Grid reinforcements} & Network & the potential to reduce grid congestion. EEHs are efficiently integrated & \\
\rowcolor{gray_light} \multirow{-2}{*}{and congestion} & integration & into the grid by i) selecting appropriate connection points for new &  \multirow{-3}{*}{Hard} \\
\rowcolor{gray_light} \multirow{-2}{*}{management} & & investments and ii) appropriately selecting grid reinforcement projects. & \multirow{-3}{*}{constraints}\\
\hline
 \rowcolor{gray_light} & & The co-optimization of the network layout and protection strategy  & \\
  \rowcolor{gray_light} \multirow{-2.0}{*}{Protection} & \multirow{-2.0}{*}{HVDC} & must consider operational security constraints,  & \multirow{-2.0}{*}{Hard} \\
  \rowcolor{gray_light} \multirow{-2.0}{*}{strategies} & \multirow{-2.0}{*}{technologies} & such as the maximum loss of infeed. & \multirow{-2.0}{*}{constraints} \\
\hline
 \rowcolor{gray}    & & The market design governing an EEH determines which interconnection topologies & \\
 \rowcolor{gray} & &  and ownership arrangements are financially viable, and therefore constrains & \\
 \rowcolor{gray} & & the set of technically feasible configurations that can be considered. In particular, & \\
  \rowcolor{gray} \multirow{-2.0}{*}{Electricity} & \multirow{-2.0}{*}{Electricity} & the chosen cost-remuneration mechanism (e.g., ex-ante or ex-post cost sharing) & \multirow{-2.0}{*}{Main} \\
 \rowcolor{gray} \multirow{-2.0}{*}{market design} & \multirow{-2.0}{*}{market design} & affects the optimal capacity and number of interconnections of the EEH, & \multirow{-2.0}{*}{drivers} \\
 \rowcolor{gray} &  & and must be aligned with the regulatory framework of each connected & \\
 \rowcolor{gray} &  & control area before a final investment decision can be made.  & \\
\hline
 \rowcolor{gray}  & & The maximum dimensions of the different building blocks of &\\
 \rowcolor{gray} \multirow{-2.0}{*}{Space} &  & an EEH are used to select the best combination of & \multirow{-2.0}{*}{Main} \\
 \rowcolor{gray} \multirow{-2.0}{*}{requirements} & \multirow{-3.0}{*}{Costs} & network components for a given space or power capacity. & \multirow{-2.0}{*}{drivers}\\
\hline
 \rowcolor{gray} & & An EEH should meet its power aggregation and transmission & \\
 \rowcolor{gray} \multirow{-2}{*}{CAPEX and} & & goals at the lowest possible cost while meeting the   & \multirow{-2.0}{*}{Main}\\
   \rowcolor{gray} \multirow{-2}{*}{market fitness} & \multirow{-3.0}{*}{Costs} & existing technical and space requirements. & \multirow{-2.0}{*}{drivers} \\
\hline
  \rowcolor{gray}     & & The designed lifetime and planned maintenance actions are &\\
  \rowcolor{gray}     & & design parameters depending on the characteristics of the EEH. &\\
 \rowcolor{gray}      & & The (potential) unavailability of the project should be& \\
 \rowcolor{gray}      & & minimized by organizing effective planned maintenance actions &\\
 \rowcolor{gray} OPEX & & and investing in reliable components, especially if the EEH & \\
\rowcolor{gray}       & & is placed in harsh weather conditions. The EEH capacity &\\
\rowcolor{gray}       & & can lead to increases in the required frequency restoration reserves.&\\
\rowcolor{gray}       & & These need to be considered in the OPEX estimates of a candidate & \multirow{-8}{*}{Main}\\
\rowcolor{gray}       & \multirow{-9.0}{*}{Costs} & project and might influence the EEH's optimal capacity. & \multirow{-8}{*}{drivers}\\
\hline
 \rowcolor{gray}    Future  & Future & Expandability is one of EEHs' key purposes for the power grid &\\
 \rowcolor{gray}    proofness  & proofness & of the future. It consists of the integration of adequate provisions & \\
 \rowcolor{gray} and & and & for interface facilitation in the system design. A balance & \\
 \rowcolor{gray} modular  & modular & between the controlled costs of prior investments and those of later & \multirow{-4}{*}{Main}\\
 \rowcolor{gray}  expandability & expandability & uncertain developments needs to be found to achieve economic efficiency.  & \multirow{-4}{*}{drivers}\\
\hline
 \rowcolor{gray}   & & A sustainable EEH requires minimization of its associated CO$_{2}$ emissions,  &\\
 \rowcolor{gray} Environmental and & & in-depth comprehension and willingness to support the surrounding &\\
 \rowcolor{gray} ecological impacts & &  natural habitat, together with empathy towards the vast community of & \multirow{-3}{*}{Main}\\
 \rowcolor{gray} & \multirow{-4}{*}{Sustainability} & direct and indirect project participants and stakeholders. & \multirow{-3}{*}{drivers}\\
\hline
 \rowcolor{gray_dark} &   &  The cable configuration, MMC converter type and protection strategy choice &\\
 \rowcolor{gray_dark}  \multirow{-2}{*}{Cable configuration,}     & & influences the maximum available power transfer of a multi-terminal DC grid. &\\
 \rowcolor{gray_dark}  \multirow{-2}{*}{converter type and }    & \multirow{-3}{*}{HVDC} & The EEH's flexibility, redundancy, and extendability should be prioritized,  & \multirow{-3}{*}{Key}\\
 \rowcolor{gray_dark}  \multirow{-2}{*}{busbar arrangement}     & \multirow{-3}{*}{technologies} & but case-specific arrangements depend on the project's general goal. & \multirow{-3}{*}{considerations}\\
\hline
 \rowcolor{gray_dark}  & & The losses in the system are a design choice  &\\
 \rowcolor{gray_dark} \multirow{-2}{*}{Losses} & & for system operators. They can account for some percentage & \\
 \rowcolor{gray_dark} \multirow{-2}{*}{Requirements} & \multirow{-3}{*}{Losses} & of the total investment-related costs over the project's lifetime. & \multirow{-4}{*}{Key} \\
\hline
 \rowcolor{gray_dark} & & The availability of EEHs heavily influences their benefits  & \multirow{-4}{*}{considerations} \\
 \rowcolor{gray_dark} & \multirow{-2}{*}{Reliability} & to the socioeconomic welfare and thus needs to be maximized & \\
 \rowcolor{gray_dark} \multirow{-2}{*}{RAM analysis} & \multirow{-2}{*}{Availability} & within the boundaries of reasonable capital expenditures. It is a & \multirow{-3}{*}{Key}\\
 \rowcolor{gray_dark} & \multirow{-2}{*}{Maintainability} & natural balance between system simplicity and integrated redundancy. & \multirow{-3}{*}{considerations} \\
\hline
\end{tabular}
\caption{Classification of the identified technical design constraints and considerations for considering investments in electrical energy hubs (EEHs) and, more generally, grid expansion planning projects according to three different criticality classes, namely \textit{hard constraints}, \textit{main drivers}, and \textit{key considerations}.}
\label{tab:constraints_summary}
\end{table}

Table~\ref{tab:classification} characterizes each identified item along three further dimensions that situate it within the planning process, but without by themselves fixing its criticality. The first is whether the item is \textit{quantitatively deterministic given a grid code}, i.e. whether an applicable code translates it into a specific numerical limit. The second is whether it is \textit{area- and regulation-dependent} or instead common across jurisdictions. The third is the \textit{planning phase} from~\cite{Stu_for_plan_HVDC} where it is involved, namely the \textit{transmission expansion plan} phase (Phase~1), in which the overall capacity, topology, and operation philosophy are fixed, or the \textit{feasibility studies} phase (Phase~2), in which the design parameters of the individual network components are determined. These dimensions explain \textit{how} a given constraint is expressed and \textit{where} it is handled. For example, a hard constraint may be numerically fixed by a grid code (\textit{Losses requirements}) or, although binding for security, depend on area-specific limits (\textit{grid reinforcements and congestion management}). Some identified constraints can be both \textit{quantitatively deterministic given a grid code} and \textit{area- and regulation-dependent}, such as \textit{network operational security}, \textit{protection strategies} and \textit{space requirements}. Therefore, these dimensions are complementary to the criticality classes from Table~\ref{tab:constraints_summary}.

\begin{table}[h]
  \centering
  \begin{tabular}{|c|c|c|c|}
    \hline
    Identified technical & Quantitatively deterministic & Area- \& regulation- & Phase 1: planning vs \\
    constraint/consideration & in presence of a grid code & dependent & Phase 2: feasibility studies \\
    \hline
    Network operational security & \ding{51} & \textcolor{black}{\ding{51}} & Phase 1\\
    \hline
    Grid reinforcements and&  & & \multirow{2}{*}{Phase 1}\\
    congestion management  & \multirow{-2}{*}{\ding{51}} & \\
    \hline
    Protection strategies & \ding{51} & \textcolor{black}{\ding{51}} & Phase 1\\
    \hline
    Electricity market design &  & \ding{51}& Phase 1\\
    \hline
    Space requirements & \ding{51} & \textcolor{black}{\ding{51}} & Phase 1\\
    \hline
    CAPEX and market fitness &  & \ding{51}& Phase 1\\
    \hline
    OPEX &  & \ding{51}& Phase 2\\
    \hline
    Future proofness and &  & & \multirow{2}{*}{Phase 1}\\
    modular expandability & \multirow{-2}{*}{\ding{51}} && \\
    \hline
    Environmental and &  & & \multirow{2}{*}{Phase 1}\\
    ecological impacts &  & \multirow{-2}{*}{\ding{51}}& \\
    \hline
    Cable configuration, converter & & & \multirow{2}{*}{Phase 2}\\
    type and busbar arrangement & \multirow{-2}{*}{\ding{51}} && \\
    \hline
    Losses requirements & \ding{51} && Phase 2\\
    \hline
    RAM analysis & \ding{51} && Phase 2\\
    \hline
  \end{tabular}
  \caption{Classification of identified technical constraints and considerations for electrical energy hubs according to their quantitative determinism under grid codes and their dependence on area-specific and regulatory frameworks.}
  \label{tab:classification}
\end{table}
Three placements warrant explicit justification, as the criterion applies to them in a way that is not immediately obvious:
\begin{itemize}
    \item \textit{Protection strategies} and \textit{network operational security} are classified both as hard constraints because the choice of protection strategy (and eventual different busbar arrangements) directly determines the maximum loss of infeed (LoI) that a fault on the EEH can cause. Since the maximum LoI is the quantity bounded by the \textit{network operational security} hard constraint, the \textit{protection strategy} decision sets the value of an operational security constraint and is therefore as binding as that constraint. This makes the causal chain explicit: protection strategy~$\rightarrow$~maximum LoI~$\rightarrow$~operational security.
    \item \textit{Electricity market design} is classified as a main driver rather than a hard constraint as it does not impose a pass/fail technical limit on the operation of the system or on any EEH design parameter. Instead, it determines which interconnection topologies, capacities, and ownership arrangements are financially viable, and thereby constrains the set of configurations that a planner carries forward. This restriction of the possible configurations, together with its effect on the optimal capacity and number of interconnections, is the actionable planning input that market design provides, and it performs the role of a \textit{main driver}: it shapes the operation and investment philosophy of the EEH without binding a single design parameter. Moreover, market design is tied to the regulatory framework of the area where the EEH is built (Table~\ref{tab:classification}): in a mature offshore market, a viable cost-remuneration framework is a deal-breaker for the final investment decision. In areas without established offshore frameworks or general market designs, it remains a policy question under development.
    \item The RAM analysis is classified as a key consideration. We note that the criticality denotes the level at which an item acts in the planning hierarchy of an EEH and the degree to which it is codified as a deterministic operational security constraint, not its engineering importance. TSO connection codes can specify binding availability figures, such as minimum availability percentages or maximum planned-outage windows. In that sense, RAM is of fundamental importance for the EEH. However, RAM enters EEH planning as a component- and asset-related requirement resolved in the feasibility-study phase, after the system-level operational-security constraints. \textit{Key consideration} should therefore be read as acting on component-level design parameters and not codified as a system-level operational-security constraint, rather than as of lower importance.
\end{itemize}

\begin{figure*}
\centering
\includegraphics[width=0.85\linewidth]{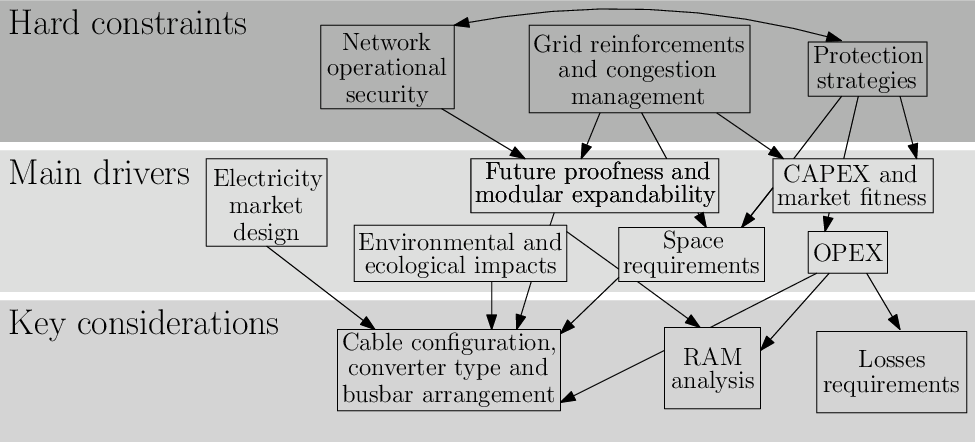}
\caption{Dependencies between the different relevant areas of interest for the identified technical constraints and considerations for the planning of investments in electrical energy hubs (EEHs) and, more generally, grid expansion projects. Based on our classification in three different criticality classes, namely \textit{hard constraints}, \textit{main drivers}, and \textit{key considerations}, the arrows link the most stringent constraints and considerations, the \textit{hard constraints}, to the \textit{main drivers}, related to the overall operation and control philosophy of an EEH. Subsequently, the \textit{main drivers} are linked to the \textit{key considerations}, related to the design parameters of the different network components.}
\label{fig:hierarchy}
\end{figure*}

Figure~\ref{fig:hierarchy} indicates the resulting hierarchy and the dependencies among the identified areas of interest. Each arrow represents a direct influence from one area of interest to the area at the other end of the arrow. The three \textit{hard constraints} ensure the operational security of the EEH design, and propagate the resulting requirements to the \textit{main drivers}. Note that \textit{protection strategies} and \textit{network operational security} are interlinked because they are interdependent, as mentioned earlier in this Section. Moreover, \textit{network operational security} and \textit{grid reinforcements and congestion management} drive \textit{future proofness and modular expandability}, while the latter also influence the \textit{space requirements} and \textit{CAPEX and market fitness} together with the \textit{protection strategies}. The \textit{protection strategies} are also related to the \textit{CAPEX and market fitness} and \textit{OPEX}. The \textit{main drivers} in turn relate to the \textit{key considerations}: \textit{electricity market design}, \textit{space requirements}, and \textit{future proofness and modular expandability} and \textit{OPEX} jointly shape the \textit{cable configuration, converter type, and busbar arrangement}, while \textit{OPEX} also influences \textit{losses requirements} and the \textit{RAM analysis}. Finally, the \textit{environmental and ecological impacts} constraints considerations on the \textit{cable configuration, converter type, and busbar arrangement} and \textit{RAM analysis}, too.

\section{Conclusion and future work} \label{sec:conclusion}
This paper aimed to identify the most relevant technical design constraints and considerations when considering investments in electrical energy hubs (EEHs), and more generally, transmission grid expansion planning projects. We identified several areas of interest for such constraints and considerations, namely \textit{network integration}, \textit{HVDC technologies}, \textit{costs}, \textit{electricity market design}, \textit{future proofness \& modular expandability}, \textit{reliability-availability-maintainability}, and \textit{sustainability}. Each area of interest is discussed separately throughout the paper. Furthermore, the relevant constraints and considerations are divided into three main criticality classes according to their importance in the EEH planning process, namely \textit{hard constraints}, \textit{main drivers}, and \textit{key considerations}. Based on this classification, we assign each identified area of interest to a class and derive the dependencies among all the areas of interest.

Future work will deal with including the identified constraints in mathematical optimization models used for evaluating the economic benefit brought by an EEH to the overall society, such as optimal power flow simulations and transmission network expansion planning problems. By including the identified constraints in such models, cost-benefit analyses taking into account the physical constraints of power systems can be proposed, and the effectiveness of grid expansion projects can be more precisely quantified. In addition, a more transparent and unified methodology for transmission grid expansion planning projects will be developed based on the findings of this paper, which could lead to reduced delays in the planning and construction of new projects, reducing the renewable energy curtailment due to such delays.


\clearpage
\bibliographystyle{ieeetr}
\bibliography{Bibliography}




\end{document}